\documentclass[12pt]{article}

\usepackage{amssymb,amsmath} 
\usepackage{amsfonts}
\usepackage[body={17cm,23.5cm}]{geometry}
\begin{document}

\title{\bf  The Weight of Vacuum Fluctuations}
\author{ \bf{Eduard Mass{\'o}}\\
\small{ \it Grup de F{\'\i}sica Te{\`o}rica and Institut de F{\'\i}sica d'Altes Energies,}\\
\small{  \it Universitat Aut{\`o}noma de Barcelona, 08193 Bellaterra, Spain}}

\date{\small{Preprint UAB-FT-663}}

\maketitle

\centerline{ \large \bf Abstract}

We examine the gravitational properties of Lamb shift energies.
Using available experimental data we show that these energies have a standard  gravitational behavior at the level of $\sim 10^{-5}$. 
We are motivated by the point of view that Lamb shift energies may be interpreted as a consequence of vacuum fluctuations of the electromagnetic field.
If this is the case, our result is a test of
the gravitational properties of quantum fluctuations.
The result is of interest in relation to the problem of the zero-point energy contribution to the cosmological constant. Indeed, the problem presupposes that the zero-point energy gravitates as all other forms of energy, and this supposition is what we test.

 \begin{center}
---------------------------------------------------------------------
 \end{center}
 


\section{The Cosmological Constant and Vacuum Fluctuations}

Present cosmological data are well fitted by assuming the existence of a dark energy component that induces acceleration of the universe.
This component is compatible with a simple cosmological constant term in the Friedmann equation, with an energy density
\begin{equation}
 \rho \simeq ( 2 \times 10^{-3} \ {\rm eV})^4
\label{rho}
\end{equation} 
in natural units, and pressure $p=-\rho$. With a few more parameters determined by data we have a sound description of all observations in the framework of the Standard Cosmological Model \cite{Frieman:2008sn}.

The Standard Model of particles and their gauge electroweak and color interactions 
is also a well established model, able to account for all experimental particle physics data. In this model there are contributions 
to the cosmological constant. This is due to the fact that the model is based on quantum fields and, in vacuum, fields fluctuate around their minimal values giving a contribution to the vacuum energy. Provided we only measure energy differences, we can subtract this type of contributions and we do not need to worry when performing calculations (technically we call it normal ordering). However, this procedure is no longer possible in the presence of gravitation for in this case the absolute value of energy matters. One expects then a net cosmological constant from the zero-point field fluctuations \cite{Zel'dovich:1968zz}. 
It has been known for many years that these contributions exceed the observed value 
(\ref{rho}) by many orders of magnitude. To solve this problem is one of the present challenges for Physics
\cite{Weinberg:1988cp}. The problem presupposes that vacuum fluctuations have the same gravitational properties as all other forms of matter. While there is no a priori reason to doubt of such presupposition, it would be desirable to test it experimentally.
Such a endeavor might seem fanciful, but in this Letter we will see that it could be possible.

It is often argued that experimental support for the presence of electromagnetic vacuum fluctuations is provided by the  verification of the Casimir effect \cite{Casimir:1948dh} in the laboratory \cite{Lamoreaux:2005gf}. 
Indeed, the Casimir force can be deduced by considering the difference among the contribution to vacuum energies in two physical situations. It follows that it is not the absolute value of the vacuum energy what we measure but energy differences. Even so, it is interesting to study the gravitational properties of the Casimir energy. This has been done theoretically in \cite{Fulling:2007xa}, with the conclusion that the Casimir energy gravitates according to the Equivalence Principle.

The measure and the theoretical prediction of the Lamb shift (LS) energy is historically considered a milestone in quantum field theory \cite{Kinoshita}.  
Along with the Casimir force, the LS energy can also be considered as a test of the vacuum fluctuations
of the electromagnetic field. While perhaps this is not very obvious in Bethe's original treatment \cite{Bethe}, it is clear in the calculation by Welton \cite{Welton:1948zz}. He deduced the LS formula corresponding to the main contribution -see below, equation (\ref{LS})- taking into consideration the effect of electromagnetic field fluctuations on the electron.
Perhaps it is even more clear in the treatment given by Power  \cite{Power}, who followed suggestions by Feynman. In \cite{Power} the deduction of the LS is based on the consideration of the change in the zero-point field energy in a box containing atoms compared to the same box without atoms, and again one finds exactly the usual formula for the LS. 
The deduction in  \cite{Power} has a striking similarity with the usual zero-point energy deduction of the Casimir effect. There are quantum field textbooks \cite{bdiz} reviewing all these topics. In the textbook \cite{Milonni} we can find a study with special emphasis in the connection with vacuum fluctuations.

The point of view that the Casimir effect is unambigous evidence for vacuum fluctuations has been disputed by Jaffe \cite{Jaffe:2005vp}.  Indeed, he has calculated the effect with no reference whatsoever to vacuum. Although we are not aware of a similar calculation for the LS, a judicious attitude should be the following.
Even if the LS can be calculated as originating from vacuum fluctuations, 
 to take the point of view that the LS is evidence for quantum fluctuations has to be considered an assumption. Once this assumption is made, it is clearly worth to test the gravitational properties of the LS energy. In the next section we show how to test the Equivalence Principle (EP) for the LS energy.

\section{Lamb Shift Energy and the Equivalence Principle}

The EP is a fundamental postulate in gravitational physics which has been experimentally tested to a high precision. We wish to test the principle for LS energies, or in other words, we would like to know to which extent this energy couples to external gravitational physics as all other types of matter and energy.

In order to proceed, we allow for a violation of the EP. We write the following relation between the gravitational mass $m_{gr}$ of an object and its inertial mass $m_{in}$,
\begin{equation}
\label{g_i}
m_{gr}=m_{in} - \lambda\, \Delta E_{LS}
\end{equation}
Here  $ \Delta E_{LS}$ is the total LS energy contributions to the object 
and $\lambda$ is a parameter that would signal a violation of the EP:
$\lambda=0$ corresponds to the case that the EP is obeyed, and $\lambda=1$ would correspond to a situation where the LS energies do no gravitate at all. Our main purpose is to get a bound on $\lambda$.

The main contribution to the LS energy of an electron in a level of principal quantum number $n$ in the H atom is given by
\begin{equation}
\label{LS}
\Delta E_{LS} (H) =   \frac{4 \alpha^2}{3} \frac{1}{m_e^2}\  \left(\log  \frac{m_e}{\epsilon_n(H)} \right) | \Psi_n (0)|^2
\end{equation}
Here, $\alpha$ is the fine structure constant, $m_e$ is the electron mass, and $\Psi_n(0)$ is the electron wave function at the origin. The parameter $\epsilon_n(H)$ is an average excitation energy, which has to be calculated numerically. For example, for the $2s$ level, $m_e/\epsilon\simeq 2 \times 10^3$.
The 2s$_{1/2}$-2p$_{1/2}$ LS in H is very well measured, and one needs to consider subdominant contributions to the energy shift apart from the main contribution (\ref{LS});
see ref. \cite{Kinoshita}.

Actually, if we apply our EP test to the atomic LS we obtain poor limits on $\lambda$. To get an interesting bound, 
we will consider the LS  of the proton energies due to the nuclear electromagnetic field.
Of course this has not been measured, but we expect a tiny contribution to the total nuclear mass coming from it. We need to calculate a nuclear effect and therefore we know in advance that some approximations have to be done due to the intrincancies of nuclear physics. 

In the nuclear shell model, the LS of the protons in a nucleus $N$ is
\begin{eqnarray}
\Delta E_{LS} (N)  &=& \frac{4 \alpha^2}{3} \frac{1}{m_p^2}\  \delta_N \nonumber \\
 \delta_N  &=&  \sum_{i } \, Z_N^{(i)} \,
 \left(\log  \frac{m_p}{\epsilon_i(N)} \right) | \Psi_i (0)|^2
 \label{EN}
\end{eqnarray}
Here, $m_p$ is the proton mass. To get $\Psi_i (0) \neq 0$ one has to restrict to $i$ running over s-wave shells of protons in the nucleus $N$. We approximate the Coulomb field on a proton by a central field produced by the other protons in inner shells. Thus, $Z_N^{(i)}$ is the total number of protons in shells inner than $i$.

In the shell model one uses the Woods-Saxon $V_{WS}$ potential and a spin-orbit term $V_{LS}$, and one gets a fairly good understanding of the spectrum and other nuclear properties. The potential is
\begin{eqnarray}
\label{WS}
V(r) &=&V_{WS}  + V_{LS} \nonumber \\
V_{WS}  &=& \frac{V_0}{1+\exp (r-R)/a} 
\end{eqnarray}
where $V_0 \simeq 50$ MeV, $R\simeq 1.3 A^{1/3}$ fm ($A$ is the atomic number), and $a\simeq 0.7$ fm. The spin-orbit coupling vanishes for the s-levels we are interested. No general analytical solution is obtained for the potential (\ref{WS}), and one has to resort to numerical calculations. For our purposes it is enough to consider two simpler potentials
where analytical results for the wave-function at the origin can be easily obtained.
 We start with the 3D harmonic (h) potential, $V^{\rm (h)}=(1/2)m_p\omega^2r^2 +$const, 
In fact, the s-wave functions near the origin for a harmonic potential with suitable chosen frequency $\omega$ and the numerically calculated wave functions corresponding to (\ref{WS})
are  practically indistinguishable \cite{bohr}. 

For the harmonic potential, if we numerate the s-waves with $n=0,1,\dots$
(corresponding to increasing energy) we find
\begin{equation}
\label{psi_ho}
 | \Psi_n^{\rm (h)} (0)|^2 = \left(   \frac{m_p\, \omega}{\pi} \right)^{3/2}\, \frac{(2n+1)!!}{n!}\, 2^{-n}
\end{equation}
We choose $\omega \simeq 9$ MeV, the value corresponding to $A=100$ \cite{bohr}, and put $n=1$. It gives
\begin{equation}
\label{50}
 | \Psi_1^{\rm (h)} (0)|^2 \simeq (50\, {\rm MeV})^3
\end{equation}
One can prove that the expression (\ref{psi_ho}) increases with $n$. 

Let us find now $\Psi(0)$ for another simple potential. We shall consider a 3D square well  (w) potential, which actually is the limit of $V_{WS}$ in (\ref{WS}) when $R\gg a$, which is a reasonable limit even for not too large $A$. For simplicity, we shall consider the infinite well, which is a valid approximation for the lowest lying states. In this case, we obtain
\begin{equation}
\label{psi_sw}
 | \Psi_n^{\rm (w)} (0)|^2 =   \frac{\pi}{2} \, \frac{n^2}{R^3}
\end{equation}
where $n=1,2, \dots$ is the quantum number.
Numerically, with the radius $R$ used before, we have
\begin{equation}
\label{40}
 | \Psi_n^{\rm (w)} (0)|^2 \simeq (40\, {\rm MeV})^3\, n^2 \, \left( \frac{100}{A} \right)^{1/3}
\end{equation}

Let us now calculate the effect of a violation of the EP when $\lambda$ is not null. Given two elements 1 and 2, such a violation is signaled by a nonzero value of the parameter
\begin{equation}
\label{eta}
\eta(1,2) =  \frac{m_{gr}}{m_{in}} \bigg|_{2}  - \frac{m_{gr}}{m_{in}} \bigg|_{1}
\end{equation}
With the EP-violating assumption (\ref{g_i}) we have a contribution to this parameter given by
\begin{equation}
\label{delta_m}
\eta(1,2) =  \lambda\ \frac{4 \alpha^2}{3} \frac{1}{m_p^3} 
\left(  \frac{\delta_{1}}{{\rm A}_{1}} -  \frac{\delta_{2}}{{\rm A}_{2}}\right) 
\end{equation}

To be conservative, we shall use the estimation (\ref{40}), because (\ref{50}) would lead to a stronger limit. Also, we point out that the log term in (\ref{EN}) does not make a big difference between elements and shells. We shall factorize it in
(\ref{delta_m}), and since we expect it to be larger than $O(1)$, we conservatively set
$\log m_p/\epsilon=1$.

Let us consider aluminium ($Z=13$, $A=27$)  and platinum ($Z=78$, $A=195$), the elements used in the Braginsky and Panov experiment \cite{Braginsky}. There are 2 s-shell protons in  aluminium,
\begin{equation}
\label{Al}
{\rm Al} = \{ (1\ s_{1/2})^{2p} \} + \dots
\end{equation}
We employ the usual notation with the number and type of nucleons as a superindex, and the dots indicate protons in shells other than s and neutrons.  There are 6 s-shell protons in platinum, 
\begin{equation}
\label{ }
{\rm Pt} ={\rm Al} +  \{ (2\ s_{1/2})^{2p}, (3\ s_{1/2})^{2p} \} + \dots
\end{equation}

We see that there are no inner protons for the 1s protons of aluminium. Thus, we get $\delta_{\rm Al}=0$. For platinum, for the 2s protons we shall consider the contribution of 1s protons, and for the 3s protons, we shall take into account the 1s and 2s protons. There are other protons contributing to the Coulomb field, but exactly how much is the contribution depends on the details of the potential. By ignoring them we get a conservative bound.

Numerically we have a contribution to (\ref{delta_m}) of about $\lambda \times 4 \times10^{-7}$.
Using the experimental bound \cite{Braginsky}
\begin{equation}
\label{EP_exp}
 |\eta({\rm Al},{\rm Au}) | \lesssim 10^{-12}
\end{equation}
  we are led to 
\begin{equation}
\label{result}
\lambda \lesssim 3 \times 10^{-6}
\end{equation}
Other experiments, done in laboratory or coming from lunar laser ranging data - see \cite{Fischbach} for a review- would give similar bounds. What can be done in the future in order to improve the limit (\ref{result}) ? An obvious way is to get an experimental limit better than (\ref{EP_exp}). One could also think of using other elements than Al and Pt such that the nuclear shell structure would enhance our hypothetical EP-violating effects. However, this does not work because there is no known element with $4s$ protons. The best we can do is  to choose a heavy element with $3s$ protons, which means $Z\geqslant 70$ (like Pt), and a lighter element with $1s$ protons and no other $s$-protons, namely, with $Z < 14$ (like Al). 

We should mention that the modes our test probe are basically of wavelengths $\lesssim 2R \sim 10$ fm and thus of energies $\gtrsim 2\pi/R \sim 100$ MeV.

\section{Lamb Shift as Active Gravitational Mass}

We can go a step further in the test of the gravitational properties of LS energies.
One may distinguish the active gravitational mass $M_{gr}$ of an object, which is the source of gravitational fields,  from the passive gravitational mass $m_{gr}$, which determines the force upon the object when placed in an external gravitational field \cite{Will}. 
Given two bodies, now the relevant parameter is
\begin{equation}
\label{sigma}
\sigma(1,2) =  \frac{m_{gr}}{M_{gr}} \bigg|_{2}  - \frac{m_{gr}}{M_{gr}} \bigg|_{1}
\end{equation}

Our test in the last section referred to the equality of inertial mass and passive gravitational mass for LS contributions. Now we would like to test the equality of the active and passive gravitational mass for LS contributions. Here laboratory experiments give poor limits and, fortunately, the authors of ref.  \cite{Bartlett:1986zz} have been able to show how to place very tight limits to $\sigma$ using lunar laser ranging data. They find a limit concerning  aluminium and iron,
 \begin{equation}
\label{Al_Fe}
 |\sigma({\rm Al},{\rm Fe}) | \lesssim 4 \times 10^{-12}
\end{equation}

In analogy with what we did in the last section, we allow for a difference in passive and active masses coming from an anomalous gravitational behavior of the contribution of Lamb shift energy,
\begin{equation}
\label{a_p}
m_{gr}=M_{gr} - \lambda' \, \Delta E_{LS}
\end{equation}
where now $\lambda' \neq 0$ would signal a violation of the equality among gravitational active and passive mass.
To calculate the contribution to $\sigma$ coming from a potential non-vanishing $\lambda'$ we need the structure
of s-wave protons in iron,
\begin{equation}
\label{ }
{\rm Fe} ={\rm Al} +  \{ (2\ s_{1/2})^{2p} \} + \dots
\end{equation}
Proceeding as before, we find a contribution of about $\lambda' \times 10^{-7}$ so that using (\ref{Al_Fe}) we get the limit
\begin{equation}
\label{result2}
\lambda' \lesssim 4 \times 10^{-5}
\end{equation}

In \cite{Nordtvedt} it is shown that the limit on violations of active-passive gravitational masses can be improved with respect the limits presented in \cite{Bartlett:1986zz} by a factor $\sim 40$. However, there is not the extraction to elements as in equation (\ref{Al_Fe}) and because of this we do not employ \cite{Nordtvedt}. However, it means that 
our  limit  (\ref{result2}) may be improved.

\section{Conclusion}

Given the contribution of the zero-point field fluctuations to the cosmological constant, 
it is interesting to try to know whether fluctuating fields gravitate at all, or do so anomalously.
With this motivation,
we have worked out a test of the gravitational properties of the LS energies. We have shown that they obey the EP at the level of 
$3 \times 10^{-6}$. We have also established a limit on a violation of the equality of active and passive gravitational mass for the LS contributions; the limit is $4 \times 10^{-5}$.
If we assume that LS energies are a consequence of electromagnetic vacuum fluctuations then what we are testing is the gravitational properties of these fluctuations. Our result might be of interest in such a case.

\section*{Acknowledgments}
 
 I thank Enrique Alvarez and Matthias Jamin for bringing references \cite{Nordtvedt} and 
 \cite{Milonni}, respectively, to my attention. I also thank Subhendra Mohanty for discussions.
 I acknowledge support by the CICYT Research Project FPA 2008-01430 and the
\textit{Departament d'Universitats, Recerca i Societat de la
Informaci{\'o}} (DURSI), Project 2005SGR00916.
This work was supported (in part) by the European Union through the Marie
Curie Research and Training Network "UniverseNet" (MRTN-CT-2006-035863)."

\end{document}